\documentclass[showpacs,twocolumn]{revtex4}
\usepackage{amsmath,amssymb,mathrsfs}
\usepackage{latexsym}
\usepackage{graphicx,psfrag} 

\newcommand{\bs}[1]{\boldsymbol{#1}}

\newcommand{\half}{$\frac{1}{2}$ }

\newcommand{\ket}[1]{\left|#1\right\rangle}
\newcommand{\bra}[1]{\left\langle#1\right|}

\newcommand{\braket}[2]{\bigl\langle#1\bigl|\bigr.#2\bigr\rangle}

\newcommand{\dw}{\downarrow}
\def\bd{\begin{displaymath}}
\def\ed{\end{displaymath}}
\def\be{\begin{equation}}
\def\ee{\end{equation}}
\def\bea{\begin{eqnarray}}
\def\eea{\end{eqnarray}}
\def\bi{\begin{itemize}}
\def\ei{\end{itemize}}
\def\bn{\begin{enumerate}}
\def\en{\end{enumerate}}

\def\ie{{\it i.e.},\ }

\def\ea{{\it et al.}}

\def\bglprl{BernevigGiulianoLaughlin01prl}
\def\bglprb{BernevigGiulianoLaughlin01prb}
\def\bglprlh{BernevigGiulianoLaughlin01prl2}
\def\bglprbh{BernevigGiulianoLaughlin02}
\def\bglpro{BernevigGiulianoLaughlin03}

\allowdisplaybreaks[1]

\begin{document}
\title{Is there an attraction between spinons in the Haldane--Shastry model?}
\author{Martin Greiter and Dirk Schuricht}
\affiliation{Institut f\"ur Theorie der Kondensierten Materie,\\
  Universit\"at Karlsruhe, Postfach 6980, D-76128 Karlsruhe}
\pagestyle{plain}
\begin{abstract}
  While the Bethe Ansatz solution of the Haldane--Shastry model
  appears to suggest that the spinons represent a free gas of
  half-fermions, Bernevig, Giuliano, and Laughlin
  (BGL)~\cite{BernevigGiulianoLaughlin01prl,BernevigGiulianoLaughlin01prb}
  have concluded recently that there is an attractive interaction
  between spinons.  We argue that the dressed scattering matrix
  obtained with the asymptotic Bethe Ansatz is to be interpreted as
  the true and physical scattering matrix of the excitations, and
  hence, that the result by BGL is inconsistent with an earlier result
  by Essler~\cite{Essler95}.  We critically re-examine the analysis of
  BGL, and conclude that there is no interaction between spinons or
  spinons and holons in the Haldane--Shastry model.
\end{abstract}
\maketitle

The Haldane--Shastry model
(HSM)~\cite{Haldane88,Shastry88,Haldane91prl1,Shastry92,HaldaneHaTalstraBernardPasquier92} 
plays a unique role among the integrable models of spin
$S=\frac{1}{2}$ chains.  In what might be refered to as a brilliant
theoretical coup, Haldane and Shastry discovered independently in 1988
that a trial wave function proposed by Gutzwiller~\cite{Gutzwiller63}
in 1963 provides the exact ground state to a Heisenberg type spin
Hamiltonian whose interaction strength falls off as the inverse square
of the distance between two spins on the chain.  If one imposes
periodic boundary conditions (PBCs), and embeds the one-dimensional chain
into a two-dimensional complex plane by mapping it onto the unit circle
with the $S=\frac{1}{2}$ spins located at complex positions
$\eta_\alpha=\exp\!\left(i\frac{2\pi}{N}\alpha\right)$, where $N$ denotes the
number of sites and $\alpha=1,\ldots,N$, the Hamiltonian
\begin{equation}
  \label{eq:hsham}
  H_{\text{HS}}
  =J\left(\frac{2\pi}{N}\right)^{\!2}
  \sum^N_{\alpha<\beta}\frac{\vec{S}_{\alpha}\cdot
    \vec{S}_{\beta}}{\vert \eta_{\alpha}-\eta_{\beta}\vert^2}
\end{equation}
possesses the exact ground state
\begin{equation}
  \label{eq:hsgs}
  \Psi_0(z_1,\dots,z_M)
  =\prod_{i<j}^M (z_i-z_j)^2\prod_{j=1}^M z_j,
\end{equation} 
for $N$ even, $M=\frac{N}{2}$. The corresponding state vector is given by
\begin{equation}
  \label{eq:hspace}
  \ket{\Psi_0}=\sum_{\{z_1,\dots,z_M\}}
  \Psi_0(z_1,\dots,z_M)\, 
  S_{z_1}^+\cdots S_{z_M}^+\ket{\dw\dw\dots\dw},
\end{equation}
where the sum extends over all possible ways to distribute the positions
$z_i$ of the up spins over the $N$ sites.
The model is fully integrable even for a finite number of sites; the
algebra of the (infinite) number of conserved quantities is generated
by the total spin and rapidity operators
\begin{equation}
  \label{eq:stotlambda}
  \bs{S}=\sum_{\alpha=1}^N \bs{S}_\alpha,\ \
  \bs{\Lambda}=\frac{i}{2}\sum_{\alpha\neq\beta}^N\,
  \frac{\eta_\alpha + \eta_\beta}{\eta_\alpha - \eta_\beta}\,
  (\bs{S}_\alpha\times\bs{S}_\beta)
\end{equation}
which both commute with the Hamiltonian but do not commute mutually.
The unique feature of the model, from a practical point of view, is
that in addition to its amenability to solution by the asymptotic
Bethe Ansatz
(ABA)~\cite{Sutherland,Haldane91prl1,Kawakami,HaHaldane93},
the ground state and many of the excited states (in principle, all the
ones where the spins of the spinon excitations are fully polarized)
can be written down in closed form, \ie the wave functions are known
explicitly.  In particular, the wave function for an individual spinon
excitation, which carries spin \half but no charge, at site
$\eta_\alpha$ is constructed in complete analogy to the wave function
for a quasihole excitation in a fractionally quantized Hall
liquid~\cite{Laughlin83}:
\begin{equation}
  \label{eq:spinon}
  \Psi_\alpha(z_1,\dots,z_M)=\prod_{j=1}^{M}(\eta_{\alpha}-z_{j})\,
  \prod_{i<j}^M (z_i-z_j)^2\prod_{j=1}^M z_j
\end{equation}
where $N$ odd, $M=\frac{N-1}{2}$. The model may hence be used to
illustrate the sense in which spinons are fractionally quantized
excitations: the spin of the spinon is one-half, while the Hilbert
space (\ref{eq:hspace}) is built up from spin-flips, which carry spin
one.

On a more profound level, the model is unique in that there is no spin
exchange between spinon excitations~\cite{Haldane91prl1}, which
follows directly from the commutativity of $\bs{\Lambda}$ with
$H_{\text{HS}}$.  Furthermore, the spinon excitations of the model
have or had been considered to constitute an ideal gas of
half-fermions, that is, an ideal gas of particles obeying fractional
statistics~\cite{Wilczek,Haldane91prl2}.  This view has received
strong support from Essler~\cite{Essler95}, who calculated the dressed
scattering matrix of the spinon excitations using the ABA, and found
it to be $S=\pm i$.  The fact that $S$ does not depend on the spinon
momenta implies that they are non-interacting or free; the phase $i$
implies that they obey half-fermion statistics.  This picture, and in
particular the applicability of the ABA to the HSM, were commonly
accepted until a few years ago.

In 2001, this picture was challenged by Bernevig, Giuliano, and
Laughlin (BGL)~\cite{\bglprl,\bglprb}, who investigated the nature of
the spinon interaction by working out the wave functions for the
spin-polarized two-spinon eigenstates explicitly.  They found ``clear
evidence for a short-range, attractive interaction between
spinons''~\cite{\bglprl}.  Furthermore, they ``prove rigorously that
this enhancement''---meaning a probability enhancement as the spinons
are close together---``is responsible for the square-root singularity
in the dynamical spin susceptibility''~\cite{\bglprl}, which has been
evaluated exactly in the thermodynamic limit for the HSM by Haldane
and Zirnbauer~\cite{HaldaneZirnbauer93}, and experimentally observed in
KCuF$_3$ by Tennant \ea~\cite{TennantCowleyNaglerTsvelik95}.
According to BGL, ``the experiments provide evidence that spinons do
interact and that the spinon interaction is what determines the
peculiar low-energy physics of spin-\half antiferromagnetic
chains''~\cite{\bglprb}.
BGL attribute the apparent contradiction between their results and the
ABA result to the fact that ``the interaction between spinons is
encoded in the definition of the pseudomomenta'' which label the Bethe
Ansatz solutions~\cite{\bglprl}.  In other words, they assert that it
is a special feature of the ABA technique that the spinon excitations
appear to be free, while there is in fact an attractive interaction
between them.

This line of reasoning may sound convincing at first sight.  Indeed,
in models like the Calogero--Sutherland~\cite{Calogero,Sutherland} or
the Haldane--Shastry model, the long-range interaction of the
particles or spins, respectively, is encoded in the definition of the
pseudomomenta.  The interacting degrees of freedom the Hilbert space
is built up from, the particles or spin-flips, are mapped through a
non-local and highly non-trivial transformation into a new set of
degrees of freedom, the pseudomomenta, which do not interact.  In a
sense, in the HSM both the $1/r^2$ tail of the spin-flip terms
$S_\alpha^+S_\beta^-$ as well as the ``potential energy'' term
$S_\alpha^zS_\beta^z$ are encoded in the pseudomomenta.
In the framework of the ABA, spinon excitations for the HSM correspond
to fractional holes in the otherwise uniform distribution of
pseudomomenta.  Specifically, a pair of spinons is constructed by
shifting the pseudomomenta quantum numbers $I_i$ from integer to
half-integer values or vice versa, and leaving the $I_i$'s or
pseudomomenta associated with the spinons unoccupied.  The energy of
the state is given by a sum of ``kinetic energies'' of each occupied
pseudomomentum, without an interaction between them.  The information
regarding the $1/r^2$ interaction between the original spins is no
longer accessible in this framework.

What is still accessible, however, is the information regarding the
energies of and the interaction between the spinon excitations.  The
energies of the spinons are given by the change in the kinetic
energies associated with the occupied pseudomomenta as we shift them.
The interaction between the spinons is encoded in the way this shift
in the pseudomomenta induced by one spinon is affected by the
existence of another.  In the spin one-half Heisenberg chain, for
example, there is a rather complicated change or ``screening'' of the
pseudomomenta due to an interaction between the spinons.  In the HSM,
by contrast, the creation of a spinon only induces a constant shift of
the pseudomomenta, which implies that the spinons are free.  The most
reliable way to extract this information, however, is to calculate the
spinon-spinon scattering matrix.  If the ABA is applicable to the HSM
at all, which is not garanteed {\it a priori} as the spin-spin
interaction is long-ranged, the result by Essler quoted above
unambigously confirms that the spinons are free.

In the remainder of this Letter, we resolve the contradiction between
the conclusions reached by Essler~\cite{Essler95} and
BGL~\cite{\bglprl,\bglprb}.  The result is that we find no reason to
doubt the applicability of the ABA, and completely agree with Essler's
conclusions.  We also agree with the explicit calculations of BGL, but
do not agree with their interpretation of the calculations.  In
particular, their conclusion that there is an attraction between
spinons or spinons and holons in the HSM, title to several
publications~\cite{\bglprl,\bglprb,\bglprlh,\bglprbh,\bglpro}, is
incorrect.

To begin with, it is worth noting that there is a physical reason to
be suspicious of BGL's result.  They conclude that there is an
attractive interaction between spinons, but no bound state.  If there
was an arbitrarily weak attraction, however, it would presumably yield
a bound state due to the Cooper instability~\cite{Cooper56}.  Cooper's
argument was originally formulated for two electrons outside a
completely occupied Fermi sphere, which are subject to an arbitrarily
weak attraction.  The argument is independent of the number of
dimensions.  The Fermi surface is only relevant in that it blocks
certain states, and renders the density of states available to the two
electrons at the point where their kinetic energy is minimal (\ie at
the Fermi surface) finite.
The Fermi statistics of the electrons accounts for the formation of a
spin singlet, but is not essential to the instability; for example,
one would also find a bound state if one were to use spinless bosons
instead.  The only subtlety involved in applying the argument to
spinons in the HSM is the half-fermi statistics of the spinons.  It is
not plausible to us, however, that this statistical interaction would
preclude the pairing, as there is not even an angular momentum barrier
associated with the statistical interaction in one dimension.

Let us now critically re-examine the arguements presented by BGL.  We
begin with a review of their analysis, and then explain why we
disagree.  

BGL construct exact two-spinon eigenstates for the HSM starting from
basis states with the two spinons localized at sites $\eta_\alpha$ and
$\eta_\beta$, 
\begin{equation}
  \label{eq:psiab}
  \Psi_{\alpha\beta}(z_1,\dots,z_M)
  =\prod_{j=1}^M(\eta_{\alpha}-z_{j})(\eta_{\beta}-z_{j})\,
  \!\prod_{i<j}^M (z_i-z_j)^2\!\prod_{j=1}^M z_j,
\end{equation}
where $M=\frac{N-2}{2}$ denotes the number of up or down spins
condensed in the uniform singlet sea.  The momentum space basis states
are obtained by Fourier transformation,
\begin{equation}
  \label{eq:psinm}
  \Psi_{mn}(z_1,\dots,z_M)=\sum_{\alpha,\beta}^N 
  \frac{(\bar\eta_\alpha)^m}{N}\frac{(\bar\eta_\beta)^n}{N}\,
  \Psi_{\alpha\beta}(z_1,\dots,z_M),
\end{equation}
where $M\ge m\ge n\ge 0$.  For $m$ or $n$ outside this range,
$\Psi_{mn}$ will vanish identically, reflecting the overcompleteness
of the position space basis (\ref{eq:psiab}).  Acting with the
Haldane--Shastry Hamiltonian on (\ref{eq:psinm}) yields
\begin{equation}
  \label{eq:schr1}
  H_{\text{HS}}\ket{\Psi_{mn}}=
  E_{mn}\ket{\Psi_{mn}}+\sum_{l=1}^{l_M}V_l^{mn}\ket{\Psi_{m+l,n-l}}
\end{equation}
where
\begin{eqnarray}
  \label{eq:enm}
  E_{mn}\!\!&=&\!\!
  -J\frac{\pi^2}{24}\biggl(N\!-\!\frac{19}{N}\!+\!\frac{24}{N^2}\biggr)
  \!+\!\frac{J}{2}\biggl(\frac{2\pi}{N}\biggr)^{\!\!2}
   \\[3pt] \nonumber
  &&\!\!\cdot \biggl[m\biggl(\frac{N}{2}\!-\!1\!-\! m\biggr)\biggr.
  +\,n\biggl(\frac{N}{2}\!-\!1\!-\!n\biggr)\!
  -\!\frac{m\!-\!n}{2}\biggl]\biggr.,
\end{eqnarray}
$l_M\!=\!\min(M\!-\!m,n)$, and
$V_l^{mn}\!=\!-\frac{J}{2}\left(\frac{2\pi}{N}\right)^2(m\!-\!n\!+\!2l)$.
Since the ``scattering'' of $H_{\text{HS}}$ acting on the
non-ortho\-gonal basis states $\ket{\Psi_{mn}}$ only occurs in one
direction, increasing the difference $m-n$ while keeping the ``total
momentum'' $m+n$ fixed, the (unnormalized) eigenstates of
$H_{\text{HS}}$ have energy eigenvalues $E_{mn}$ and are of the form
\begin{equation}
  \label{eq:phi}
  \ket{\Phi_{mn}}=\sum_{l=0}^{l_M} a_l^{mn}\ket{\Psi_{m+l,n-l}}
\end{equation}
with $a_0^{mn}\!=\!1$.  A recursion relation for the
coefficients $a_l^{mn}$ is easily obtained from (\ref{eq:schr1}).
Combining (\ref{eq:phi}) and (\ref{eq:psinm}), one thus obtains an
expansion of the exact energy eigenstates $\ket{\Phi_{mn}}$ in terms
of localized spinon states $\ket{\Psi_{\alpha\beta}}$.

In a technically truly remarkable analysis, BGL have further succeeded
in obtaining the coefficients $p_{mn}(\eta_{\alpha-\beta})$ in 
the inverse expansion 
\begin{equation}
  \label{eq:inv}
  \ket{\Psi_{\alpha\beta}}=
  \sum_{m=0}^M\sum_{n=0}^m(-1)^{m+n}
  \;\! \eta_\alpha^m \;\!\eta_\beta^n\;
  p_{mn}(\eta_{\alpha-\beta})\ket{\Phi_{mn}} 
\end{equation}
of the localized spinon states in terms of the energy eigenstates by
solving a hypergeometric differential equation.  Since a spin-flip 
$S_\alpha^-$ acting on the Haldane--Shastry ground
state yields a state with a pair of spinons localized at
$\eta_\alpha$,
$
  S_\alpha^-\ket{\Psi_0}=\eta_\alpha\ket{\Psi_{\alpha\alpha}},
$
the expansion of $S_q^-\ket{\Psi_0}$ in terms of $\ket{\Phi_{mn}}$
is determined by $p_{mn}(1)$,
\begin{eqnarray}
  \label{eq:spinex}
  S_q^-\!\!&\!\!&\!\!\ket{\Psi_0}
    =\sum_{\alpha=1}^N(\eta_\alpha)^k S_{\alpha}^-\ket{\Psi_0}
  \nonumber\\ 
  &\!\!&\!\!\!\!=N\!\sum_{m=0}^M\sum_{n=0}^m  
  (-1)^{m+n}\;\!\delta_{m+n+k+1,0}\,p_{mn}(1)\ket{\Phi_{mn}},\qquad
\end{eqnarray}
where $q=\frac{2\pi k}{N}$ and the Kronecker-$\delta$ is defined modulo $N$.
The explicit expression for $p_{mn}(1)$ enabled BGL to
calculate the dynamical spin susceptibility (DSS)
\begin{equation}
  \label{eq:dds}
  \chi_q(\omega)\equiv-
  \text{Im}\bra{\Psi_0}S_{-q}^+\,
  \frac{1}{\omega-(H_{\text{HS}}\!-\!E_0)+i0}\,
  S_{q}^-\ket{\Psi_0},
\end{equation}
for finite chains as well as in the thermodynamic limit, thus
providing an alternative derivation of the Haldane--Zirnbauer 
formula~\cite{HaldaneZirnbauer93}.
The DSS shows a square-root singularity at the lower threshold
frequency of the two-spinon continuum, which is a characteristic
feature of spin-\half chains~\cite{MuellerThomasBeckBonner81}.

As already mentioned, we completely agree with these calculations.
We disagree, however, with BGL's interpretation of the results as
evidence for a spinon attraction.  

The first argument given by BGL in favor of a spinon interaction is
based on a plot of $|p_{mn}(e^{i\theta})|^2$ for $m=M$, $n=0$, as a
function of $\theta$.  They interpret $|p_{mn}(e^{i\theta})|^2$ as
probability for finding the spinons at a distance $\theta$ along the
circle from each other, and show it to be strongly enhanced at small
$\theta$.  
The problem with the argument is that, as one can easily see from
(\ref{eq:inv}), the $p_{mn}(\eta_{\alpha-\beta})$'s are the
coefficients in the expansion of the overcomplete basis states
$\ket{\Psi_{\alpha\beta}}$ at fixed $\alpha$, $\beta$ in terms of
$\ket{\Phi_{mn}}$.  Due to this overcompleteness,
the $p_{mn}(\eta_{\alpha-\beta})$'s as functions of $\eta_{\alpha-\beta}$
have no direct physical interpretation.
The actual relative spinon-spinon wave function
$\varphi_{mn}(\eta_{\alpha-\beta})$ for given $m$ and $n$ provides the
coefficients in
\begin{equation}
  \label{eq:relwave1}
  \ket{\Phi_{mn}}=\sum_{\alpha=1}^N\sum_{\beta=1}^N
  \,\varphi_{mn}^{\phantom{\dagger}}(\eta_{\alpha-\beta})
  \cdot(\eta_{\alpha+\beta})^{\textstyle\frac{m+n}{2}}\,
  \frac{\ket{\Psi_{\alpha\beta}}}
  {\|\!\ket{\Psi_{\alpha\beta}}\!\|}.
\end{equation}
It is easily seen from (\ref{eq:phi}) and (\ref{eq:psinm}) that a possible
choice for $\varphi_{mn}(\eta_{\alpha-\beta})$ is
\begin{equation}
  \label{eq:relwave2}
  \varphi_{mn}^{\phantom{\dagger}}(\eta_{\alpha-\beta})=
  \sum_{l=0}^{l_M}\,a_l^{mn}\cdot
  (\eta_{\alpha-\beta})^{\textstyle\frac{m-n+2l}{2}}\,
  \|\!\ket{\Psi_{\alpha\beta}}\!\| .
\end{equation}
Depending on $m$ and $n$, one finds that $\varphi_{mn}(e^{i\theta})$
is sometimes enhanced and sometimes suppressed for small $\theta$, but
even if there was a clear enhancement, it would not allow for a
conclusion regarding a spinon attraction.  The reason is simply that
the overcompleteness of the basis states $\ket{\Psi_{\alpha\beta}}$
implies that $\varphi_{mn}(\eta_{\alpha-\beta})$ is not uniquely
determined, \ie there are infinitely many choices for
$\varphi_{mn}(\eta_{\alpha-\beta})$ which yield the same
$\ket{\Phi_{mn}}$ in (\ref{eq:relwave1}).

The second argument of BGL is that the last term in the energy
(\ref{eq:enm}) of the two-spinon state $\ket{\Phi_{mn}}$ represents
``a negative interaction contribution that becomes negligibly small in
the thermodynamic limit''~\cite{\bglprl}.  
The problem here is that BGL identify the momenta $q_m$ and $q_n$ of
the individual spinons according to
\begin{equation}
  \label{eq:bglqm}
  q_{m}=\frac{\pi}{2}-\frac{2\pi}{N}\!\left(m+\frac{1}{2}\right),\  
  q_{n}=\frac{\pi}{2}-\frac{2\pi}{N}\!\left(n+\frac{1}{2}\right),  
\end{equation}
and interpret the two preceding terms in (\ref{eq:enm}) as the
kinetic energies of the individual spinons.
The correct identification of the spinon momenta for $m\ge n$, however, is
\begin{equation}
  \label{eq:qm}
  q_m=\frac{\pi}{2}-\frac{2\pi}{N}\!\left(m+\frac{3}{4}\right),\
  q_n=\frac{\pi}{2}-\frac{2\pi}{N}\!\left(n+\frac{1}{4}\right),
\end{equation}
which implies that the kinetic energy of the spinons is 
given by all three terms in the square bracket in (\ref{eq:enm}).  With
$E(q)=\frac{J}{2}\bigl[\bigl(\frac{\pi}{2}\bigr)^2-q^2\bigr]$,
one finds
\begin{equation}
  \label{eq:etot}
  E_{mn}=-J\frac{\pi^2}{24}\!\left(N\!+\!\frac{5}{N}\!-\!\frac{6}{N^2}\right)
  +E(q_m)+E(q_n).
\end{equation}
The alleged spinon interaction term has disappeared.  Physically, the
shift between $q_m$ and $q_n$ by one-half of a momentum spacing
$\frac{2\pi}{N}$ is nothing but a manifestation of the half-fermi
statistics of the spinons.  While the allowed values for the total
momenta $q_m+q_n$ are those for PBCs, the allowed values for the
difference in the momenta $q_m-q_n$ are those for anti-PBCs, \ie PBCs with the
ring threaded by a flux $\pi$.

Finally, BGL claim to prove that the enhancement of $|p_{mn}(e^{i\theta})|^2$
they find when plotting it as a function of the spinon separation
$\theta$ is responsible for the square-root singularity in the
DSS~\cite{\bglprl}.
Their proof then consists of the derivation of the Haldane--Zirnbauer
formula sketched above.  In their longer paper~\cite{\bglprb}, BGL conclude that
their ``analysis definitely proves that the square-root sharp edge on
top of the broad spectrum is nothing but the interaction between
spinons'', and say that ``this result is of the utmost importance,
since it represents a way to experimentally test the interaction among
spinons in one dimension''.

There are several problems attached to this line of reasoning.  First,
the coefficients $|p_{mn}(e^{i\theta})|^2$ cannot be interpreted as a
probability as a function of the spinon separation $\theta$, as
explained above.
Second, it is not $p_{mn}(e^{i\theta})$ as a function of $\theta$ for
fixed $m$ and $n$ which enters the derivation of the
Haldane--Zirnbauer formula, but $p_{mn}(1)$ as a function of $m$ and
$n$, as one can directly see from (\ref{eq:spinex}).

The square-root singularity in the DSS is not due to an alleged spinon
attraction, but a general consequence of the fractional quantization
of spin excitations in spin-\half chains.  The position space basis
for these fractional excitations, the spinons, is necessarily
overcomplete.  The local creation of two spinons through a spin-flip
is not equivalent to a creation of all two-spinon energy eigenstates
with the same relative weight.  The process rather creates
predominantly spinons with lower energies, which is reflected in the
square-root singularity in the DSS.  Since the fractional quantization
of spin excitations is a generic feature of spin-\half chains, the
square-root singularity in the DSS is generic as well.  It exists in
the HSM, where spinons are free, but also in the Heisenberg model,
were spinons are interacting.  With the experimental observation of
the square-root singularity in KCuF$_3$, 
Tennant \ea~\cite{TennantCowleyNaglerTsvelik95} have observed fractional
quantization in spin chains.  They did not observe a spinon-spinon
attraction.

In the context of this analysis, it is worthwhile to mention a
curiosity of the two-spinon eigenstates.  In the evaluation reviewed
above, BGL obtained the coefficients $a_l^{mn}$ by explicitly solving
the Sutherland equation (\ref{eq:schr1}) using the Ansatz
(\ref{eq:phi}).  The coefficients $a_l^{mn}$ were hence determined by
the Hamiltonian, and appear to contain information inflicted on the
system by the Hamiltonian.  In principle, this could include information 
regarding an interaction between spinons.

In fact, however, the Hamiltonian is not even required in determining
the coefficients $a_l^{mn}$.  If we wish to construct an orthogonal
basis $\ket{\Phi_{mn}}$ according to (\ref{eq:phi}) with
$a_0^{mn}\!=\!1$ from the non-orthogonal basis $\ket{\Psi_{mn}}$, the
overlaps $\braket{\Psi_{mn}}{\Psi_{m'n'}}$ for all $m,n,m',n'$
completely determine all the coefficients $a_l^{mn}$, as the reader
will be able to verify easily for himself.  The coefficients
$a_l^{mn}$ as well as $p_{mn}(\eta_\alpha)$, and therefore also the
``scattering amplitudes'' $V_l^{mn}$ in (\ref{eq:schr1}), hence
contain no information except the one regarding the Hilbert space
structure of the fractionally quantized excitations.  Accordingly, it
seems impossible as a matter of principle to reach a conclusion
regarding an interaction between the spinons by studying these
coefficients.

In conclusion, we have shown that the spinons in the HSM represent an
ideal gas of half-fermions, and thereby dispersed all evidence that
the ABA might not be applicable to the model.  An ana\-ly\-sis similar
to the one presented here shows that there is likewise no interaction
between spinons and holons in the HSM.  The conclusions drawn by BGL
with regard to this
question~\cite{BernevigGiulianoLaughlin01prl2,BernevigGiulianoLaughlin02}
are likewise incorrect.

We wish to thank N.~Andrei for sharing his expertise on the Bethe
Ansatz with us.  This work was partially supported by the German
Research Foundation (DFG) through GK 284.


\begin{thebibliography}{10}

\bibitem{BernevigGiulianoLaughlin01prl}
B.~A. Bernevig \ea, Phys. Rev. Lett. {\bf 86},  3392  (2001).

\bibitem{BernevigGiulianoLaughlin01prb}
B.~A. Bernevig \ea,  Phys. Rev. B {\bf 64},  24425  (2001).

\bibitem{Essler95}
F.~H.~L. E{\ss}ler, Phys. Rev. B {\bf 51},  13357  (1995).

\bibitem{Haldane88}
F.~D.~M. Haldane, Phys. Rev. Lett. {\bf 60},  635  (1988).

\bibitem{Shastry88}
B.~S. Shastry, Phys. Rev. Lett. {\bf 60},  639  (1988).

\bibitem{Haldane91prl1}
F.~D.~M. Haldane, Phys. Rev. Lett. {\bf 66},  1529  (1991).

\bibitem{Shastry92}
B.~S. Shastry, Phys. Rev. Lett. {\bf 69},  164  (1992).

\bibitem{HaldaneHaTalstraBernardPasquier92}
F.~D.~M. Haldane \ea, Phys. Rev. Lett. {\bf 69},  2021  (1992).

\bibitem{Gutzwiller63}
M.~C. Gutzwiller, Phys. Rev. Lett. {\bf 10},  159  (1963).

\bibitem{Sutherland} B. Sutherland, J. Math. Phys. {\bf 12}, 246
  (1971); {\it ibid.}\  251; Phys. Rev. A {\bf 4}, 2019
  (1971); {\it ibid.}\ {\bf 5}, 1372 (1972).
  
\bibitem{Kawakami} N. Kawakami, Phys. Rev. B {\bf 45}, 7525 (1992);
  {\it ibid.}\ {\bf 46}, 1005 (1992).

\bibitem{HaHaldane93}
Z.~N.~C. Ha and F.~D.~M. Haldane, Phys. Rev. B {\bf 47},  12459  (1993).

\bibitem{Laughlin83}
R.~B. Laughlin, Phys. Rev. Lett. {\bf 50},  1395  (1983).

\bibitem{Wilczek}
F. Wilczek, {\em Fractional statistics and anyon superconductivity} (World
  Scientific, 1990).

\bibitem{Haldane91prl2}
F.~D.~M. Haldane, Phys. Rev. Lett. {\bf 67},  937  (1991).

\bibitem{HaldaneZirnbauer93}
F.~D.~M. Haldane and M.~R. Zirnbauer, Phys. Rev. Lett. {\bf 71},  4055  (1993).

\bibitem{TennantCowleyNaglerTsvelik95}
D.~A. Tennant \ea, Phys. Rev. B {\bf
  52},  13368  (1995).

\bibitem{Calogero} F. Calogero, J. Math. Phys. {\bf 10}, 2191 (1969);
  {\it ibid.}\ 2197.

\bibitem{BernevigGiulianoLaughlin01prl2}
B.~A. Bernevig \ea, Phys. Rev. Lett. {\bf 87}, 177206  (2001).

\bibitem{BernevigGiulianoLaughlin02}
B.~A. Bernevig \ea, Phys. Rev. B {\bf 65}, 195112 (2002).

\bibitem{BernevigGiulianoLaughlin03}
  B.~A. Bernevig \ea, in {\em Quantum phenomena in
  mesoscopic systems}, eds.\ B.\ Altshuler \ea\ (IOS Press, 2003).

\bibitem{Cooper56}
L.~N. Cooper, Phys. Rev. {\bf 104},  1189  (1956).

\bibitem{MuellerThomasBeckBonner81}
G. M\"uller \ea, Phys. Rev. B {\bf 24},  1429
   (1981).

\end{thebibliography}
\end{document}